\documentclass[sigconf,screen,authorversion,hyphen]{acmart}

\usepackage{breakurl}
 \hypersetup{
           breaklinks=true,   
           colorlinks=true,   
           pdfusetitle=true,  
        }

\setcopyright{rightsretained}
\acmPrice{}
\acmDOI{10.1145/3611643.3613900}
\acmYear{2023}
\copyrightyear{2023}
\acmSubmissionID{fse23industry-p105-p}
\acmISBN{979-8-4007-0327-0/23/12}
\acmConference[ESEC/FSE '23]{Proceedings of the 31st ACM Joint European Software Engineering Conference and Symposium on the Foundations of Software Engineering}{December 3--9, 2023}{San Francisco, CA, USA}
\acmBooktitle{Proceedings of the 31st ACM Joint European Software Engineering Conference and Symposium on the Foundations of Software Engineering (ESEC/FSE '23), December 3--9, 2023, San Francisco, CA, USA}
\received{2023-05-18}
\received[accepted]{2023-07-31}

\AtBeginDocument{%
  \providecommand\BibTeX{{%
    \normalfont B\kern-0.5em{\scshape i\kern-0.25em b}\kern-0.8em\TeX}}}

\begin{document}
\title[Understanding Hacker's Work]{Understanding Hackers' Work:\\ An Empirical Study of Offensive  Security Practitioners}

\author{Andreas Happe}
\email{andreas.happe@tuwien.ac.at} 
\orcid{0009-0000-2484-0109}

\affiliation{
    \institution{TU Wien}
    \city{Vienna}
    \country{Austria}
}
\author{Jürgen Cito}
\email{juergen.cito@tuwien.ac.at} 
\orcid{0000-0001-8619-1271}
\affiliation{
    \institution{TU Wien}
    \city{Vienna}
    \country{Austria}
}

\begin{abstract}
Offensive security-tests are commonly employed to pro-actively discover potential vulnerabilities. They are performed by specialists, also known as penetration-testers or white-hat hackers. The chronic lack of available white-hat hackers prevents sufficient security test coverage of software. Research into automation tries to alleviate this problem by improving the efficiency of security testing. To achieve this, researchers and tool builders need a solid understanding of how hackers work, their assumptions, and pain points.

In this paper, we present a first data-driven exploratory qualitative study of twelve security professionals, their work and problems occurring therein. We perform a thematic analysis to gain insights into the execution of security assignments, hackers' thought processes and encountered challenges. This analysis allows us to conclude with recommendations for researchers and tool builders, to increase the efficiency of their automation and identify novel areas for research.
\end{abstract}



    
    \maketitle
    
    \section{Introduction}
    
    For convenience and efficiency reasons, more and more devices are being connected and thus exposed to public networks. While beneficial, this has a dark undercurrent: the respective system's attack surface is increased and could be exploited by malicious actors. In a perfect world, all created software would be free from faults. As recent~\cite{printernightmare,zdi}, and not so recent~\cite{durumeric2014matter}, news implies, we are sadly not there yet.
    While secure software development, enabled by defensive security testing~\cite{defensive1, defensive3, defensive4}, is the long-term goal, short-term interventions are needed. In addition, there is an ever-increasing abundance of legacy software whose security needs to be verified too. A pragmatic approach is to perform security assessments, also known as penetration tests (pen-tests), to identify vulnerabilities and remediate them before they are discovered and exploited by malicious actors. 

    This approach is limited by the availability of skilled offensive security professionals~\cite{isc2, lackofsecuritypersonell}. While this situation should be remediated through increased enrollment in IT security educational programs, improving the efficiency of the penetration testers through tooling is an equally important measure. To accomplish this, research and tooling should be well-aligned with security professionals' activities and needs.
    
    However, to the best of our knowledge, there has been no empirical research into what type of security assessments are performed, what actions are regularly performed within those, or how professionals select attacks to be run against their targets. Without this, developments might be swift but misguided, and thus eventually irrelevant.
    
    \paragraph{\textbf{Research Questions \& Structure of this Work}}
    
    We used three research questions to drive the development of this work; the applied research method is described in the \textsc{Methodology} section.
    
    Our first research question was \textbf{``What do common security tests look like?''} We present the gathered information in section \textsc{Performing Security Tests}, detailing different types of assignments, their particularities, common actions performed during assignments, and the role of automation.
    
    The second research question \textbf{``How do Hackers perform their tasks?''} focused on the inner world of our participants. Education is an important part of socialization, therefore, results about this aspect is included in section \textsc{Becoming A Hacker}. In Section \textsc{How do Hackers think?} we present recurring themes detected during our analysis. We focus on thought processes during assignment execution, target and attack selection, dealing with uncertainty, and internal quality assurance.
    
    The \textsc{Discussions and Implications} Section is the response to the final \textbf{``What tedious or time-consuming areas could be improved?''} question. We grouped the identified research and development opportunities according to our target audience of researchers and tool builders.
    
    \section{Related Work}

    While there has been ample research on secure software development and defensive security testing~\cite{defensive1, defensive3, defensive4}, the focus of our study is offensive security testing. To the best of our knowledge, this is the first work that focuses on how hackers work, i.e., the context within which a security professional moves and the processes that influence their decisions during security assignments.

    Huaman et al.~\cite{huaman2021large} performed a large-scale interview study of German small-to-medium enterprises (SMEs). While SMEs are making up a third of Germany's GDP, they often lack resources for establishing an effective cyber-security posture. It analyzes their preconception with regards to cybercrime, their adoption of security measures and their experiences with attacks. In contrast to our study, this focuses upon the potential ``victims'', not upon security operators. One interesting finding was that 45.1\% of interviewed companies had a cybersecurity incident warranting manual response in the preceding 12 months --- further highlighting the need for trained personnel.
    
    Smith, Theisen and Barik~\cite{smith2020case} describe Red Teams working at Microsoft. They cover a wide range of topics including how corporate culture and red teaming interact. They also lightly touched on how people became security professionals and the interactions in their daily work. Its interviewees were recruited from within Microsoft, a single large-scale company, and thus might not reflect wider industry practices which, as referenced by the previously mentioned paper, consists to a large degree of SMEs. In contrast, this publication focuses on the execution of security assignments, highlights hacker's thought processes and details challenges in academic and automation research. Furthermore, this paper is not limited to the discipline of red-teaming.
 
    Van den Hout~\cite{phdthesis} investigated the impact of different penetration test methodologies on the quality of the tests performed but concluded that only one reviewed methodology had widespread adoption, but its recommendations for a structured approach were not taken into account. This could indicate a gap between ``real'' penetration testing and codified methodologies.

    Multiple papers describe aspects of penetration-testing without focusing on the operator's mindset or their decision processes. Munaiah et al.~\cite{munaiah2019characterizing} analyze event datasets and manually map attack patterns to \textit{MITRE ATT\&CK Enterprise}. This is used to show a-posteriori attack patterns but does not analyze how hackers select the attacks to execute. \textit{MITRE ATT\&CK} itself is a taxonomy of TTPs (Tactics, Techniques, and Procedures) and not a full attack methodology. Bhuiyan et al.~\cite{bhuiyan2020vulnerability} uses GitHub security bug reports to identify the origins of bug reports. Examples of these origins are software source code, software log files, binary files, etc. This details what data are used during reporting, but does not explain how a security professional identifies potential vulnerabilities for research in the first place, e.g., why a security professional analyzes a mentioned log file for relevant security information.

    Other papers focus on narrow sub-disciplines of hacking which cannot be projected upon the hacking industry at large. Ceccato et al.~\cite{ceccato2019understanding} describes how hackers attack protected software, i.e., how software protection mechanisms in provided binary files are analyzed through reverse engineering. Based upon the responses of our interview series, reverse-engineering is not representative for activities performed by offensive operators at large.\footnote{During the interview series, a single participant mentioned using fuzzing to hunt for vulnerabilities. They were switching to other disciplines due to the high resource and time requirements of fuzzing.}

    The PhD thesis ``How Hackers Think''~\cite{summers2015hackers} is a high-level treatise on hacker history, culture and their thought processes. It identifies multiple characteristics of hackers, e.g., being highly self-motivated and curious, being able to tolerate ambiguity, and their use of mental models and patterning. Its focus lies on a high conceptual level and does not analyze how hackers actually identify and chose vulnerabilities. Neither does the study identify how different areas of penetration-testing, e.g., OT or red-teaming, might impact a hacker's mindset.
    
    \section{Methodology}
    \label{section:methodology}
    
    This paper follows a \textit{pragmatist} approach~\cite{mackenzie2006research,saunders2013layers} combining methods from the \textit{empiricist} and \textit{summarist interpretist} traditions~\cite{guba1994competing}.
    
    We used semi-structured interviews to gather insights into hackers' work and thought processes.
    
    \begin{table}[!t]
    \caption{Participants}
    \label{table_interviewees}
    \centering
    \begin{tabular}{lll}
    \toprule
    Participant & Primary & Secondary\\
    \midrule
    Participant 1 & web & infrastructure, iso27001 \\
    Participant 2 & web & infrastructure, mobile \\
    Participant 3 & red-team & AD, OT, web \\
    Participant 4 & web & social engineering \\
    Participant 5 & red-team, IoT/OT & web, social engineering\\
    Participant 6 & web & AD, social engineering \\
    Participant 7 & infrastructure & web, tool development \\
    Participant 8 & web & infrastructure \\
    Participant 9 & infrastructure & AD \\
    Participant 10 & red-team, AD \\
    Participant 11 & OT, IoT & web \\
    Participant 12 & web & \\
    \bottomrule
    \end{tabular}
    \bigskip
    \begin{center}
    \footnotesize \textit{AD} denotes Internal network tests; \textit{web}, \textit{infrastructure} and \textit{IoT} denote pen-tests.
\end{center}
    \end{table}
    
    \textbf{Ethical Considerations.} Our institution does not have a formal IRB process but offers voluntary submission to a Pilot Research Ethics Committee. As human interviews were conducted, the committee was consulted, and topics were discussed, including ethically relevant methodological clarifications, more specifically questions related to the involvement of voluntary participants in the research, as well as mitigating the risk of contextual identification. Participants gave their informed consent before the interviews took place; all data collected were anonymized by researchers prior to analysis. All data storage and processing complied with national privacy regulations and the EU's General Data Protection Regulation (GDPR).
    
    \textbf{Recruitment.} 
    We define the target population as offensive-security practitioners that work directly with customer systems.
    Previous research has found that security professionals are reluctant to communicate with outsiders~\cite{kotulic2004there}, especially when it comes to their methodology and techniques. To counteract this, researchers reached out to public figures: the initial seed was populated by contacting security companies, finalists of public security challenges, and security conference participants. We use snowball sampling to improve the interview pool: At the end of each interview, we asked the current interviewee to connect us with other offensive security professionals. In addition, we cold-called both a hacking education YouTuber and a public hacking collective that is well known for publishing vulnerability disclosures. Both were mentioned by the participants during the interviews, both did not react to the contact attempt further enforcing the idea of a close-knit community~\cite{kotulic2004there}.
    
    We sampled new interview participants until theoretical saturation was reached, that is, no new information was obtained during the interviews. When considering theoretical saturation we differentiated between common themes and themes specific to the interviewee's specialty area. We continued interviews until neither two subsequent interviews contributed new specialty area information, nor three subsequent interviews contributed new common themes. Theoretical saturation was reached after the 12th interview which fit recommendations~\cite{guest2006many, francis2010adequate}.
    
    \textbf{Participants.} We considered participants that worked primarily in an offensive security field and excluded participants that primarily worked within social engineering or physical security. If participants were working in a hybrid field, such as reverse-engineering or source-code analysis, their primary focus had to be offensive. To gain seasoned results, we only reached out to professionals with at least four years of experience in the IT security field.
    
    To our dismay, we were not able to recruit any offensive security professionals that identified as non-male. While we come from a culture that naively prides itself on blind meritocracy~\cite{thehackermanifesto}, we found this contradiction disturbing. As we did not deem it relevant, we did not ask about our participants' religious or cultural affectivities, but in hindsight, we can assume diversity in that area.
    
    To protect the anonymity of the participants, we cannot detail their employment status, ethnicity, work experience before security work, and time of employment within the security field, etc. When excluding education and CTF-participation, participants had an average work experience of 9 years ($\mu=9.0, \sigma=6.5,median=8$).
    
    \textbf{Interview Protocol.} We conducted semi-structured interviews utilizing video conferencing software. All but two interviewees enabled both video and audio transmission. The average duration of the interview was 55 minutes. Before the interview started, the participants were informed about data processing, and their rights, and asked for their informed consent.
    
    We opened the interviews with questions about the interviewee's job description and how they acquired the needed skill set. Those were followed up by talking about the types of security assignments the participants are involved with. For 1--3 of these areas, detailed questions about particularities, procedures, automation, and problems were asked; since the questions were open-ended, the interviews branched out to subtopics organically. The interviews were closed with questions about grievances and additional thoughts related to the field of IT security.
    
    We recorded and manually transcribed all interviews. During the transcription, sensitive data was scrubbed from the interview; the transcribed interview was then submitted for confirmation to the interviewee. Scrubbed interviews were loaded into \textit{delve}~\cite{delve} for thematic analysis.
    
    \textbf{Analysis.} \textit{Reflexive Thematic Analysis}~\cite{braun2019reflecting} was chosen to perform a data-driven exploratory analysis of interview transcriptions. In summary, when performing thematic analysis, the researchers initially familiarize themselves with the data, and extracts of the data are tagged with codes. These codes are then used to create clusters that identify or construct underlying themes. Then, those themes are reviewed, defined, and named. The results of the findings are presented in Section~\ref{section:becoming}--\ref{section:mind}.
    
\textbf{Data Availability.} The data used in this study was collected through interviews with a close-knit community of ethical hackers. Deanonymization would likely not be preventable. In accordance with ethical guidelines and agreement with the interview participants, the decision was made not to release the interview data. All meta-information related to the interviews, including the interview guide and consent forms are part of our replication package.

    \textbf{Threats to Validity. } Any interview-based study faces the threat of \textit{selection bias} (\textit{internal threat}). To counteract this, we performed snowball sampling, recruited random security professionals during security conferences, and explicitly invited security professionals from different disciplines. For ethical reasons, interview participation was limited to white-hat hackers (\textit{internal threat}). According to prior analysis, the activities of black-hat hackers, e.g., Ransomware groups, can be seen as a subset of the activities performed by ethical red-teams~\cite{conti, ireland_conti} which are covered in this work.
    
    Another potential bias would be \textit{experimenter bias} (\textit{internal threat}). To reduce the risk, all the data collected was analyzed separately by the different authors, and their respective labeling results were compared for differences, ambiguities were discussed and resolved.
    
    Hacking contains multiple disciplines. Our results might only capture common themes of a subset of those (\textit{external validity}). We try to counteract this by inviting interviewees from various hacking fields, as is reflected in Table~\ref{table_interviewees}. The geographical distribution covered roughly Central Europe. Other geographic regions might be more advanced when it comes to the utilization of the different types of security assignment.
    
    \section{Becoming a Hacker}
    \label{section:becoming}

    The interview responses reveal several interesting themes regarding the path to becoming a hacker.
    
    \textbf{Academic Education.} All but one participant attended at least a single university-level class. Nine completed bachelor's degree studies in IT (or related field, such as CS), and of those, all continued to add a master's level degree. The percentage of interviewees enrolled in IT security specific programs increased from 55\% ($n=5$) for bachelor's studies to 78\% ($n=7$) for master's studies. This fits the perceived lack of IT-Security and Secure Development lectures during non IT-security centric programs, which was partially addressed by attending CTFs or enrolling for non-mandatory security classes. Classes were often taken in an extra occupational capacity. All fitting a common theme of ``\textit{fascination with IT security}'' combined with high intrinsic motivation.
    
    \textbf{Experience before IT-Security.} Having 2--3 years of non-security IT exposure before entering the IT security field was found to be advantageous. Another related recommendation was to have a broad IT security base combined with one or two specialization areas. Within our group of interviewees, the common base was web security or internal network assessments; examples of specializations were red teaming or cloud-specific knowledge.
    
    \textbf{Staying relevant.} All interviewees perceived a need for ongoing education. The ubiquitous information source was Twitter/X, followed by other online services such as YouTube channels, blog posts, Reddit, Github, or commercial online courses. In the physical world, colleagues and conferences were mentioned. The quality of online material was considered high, although one interviewee had qualms about publishing information due to potential misuse. A single participant regularly used the Darknet as a news source.
    
    \textbf{To CTF or not.} CTF attendance was a common theme. Participants saw a bidirectional information transfer: skills learned in CTFs were applicable at work and vice versa. Tasks in CTFs were considered very targeted in that they narrowly focus on a vulnerability, and solving the challenge or reading a write-up were considered efficient ways of gathering knowledge about the respective vulnerability. Specialized security practitioners, e.g., from the OT or ICS area, found CTFs to be introductory and shallow.
    
    \section{How do Hackers work?}
    \label{findings}
    \label{pentests}

    While we encountered the common muttering of ``\textit{every projects is different}'', these sections identify types of penetration tests, each with distinct requirements, strategies, and particular actions. When looking at a pen-tester's work, this is the external view, i.e., how a pen-tester's work is perceived from the outside.
    
    \subsection{Types of Security Tests and their Differences}
    \label{section:project_types}
    
    \begin{table}[!t]
    \caption{Types of Security Assessments}
    \label{table_project_types}
    \centering
    \resizebox{\columnwidth}{!}{
        \begin{tabular}{lcrr}
        \toprule
        Type & Covert & Team-Size & Effort in Days\\
        \midrule
        Vulnerability Assessment & not typical & 1 & 2-4  \\
        Penetration Test         & optional & 1-2 & 5-10 \\
        Internal Network Test    & optional & 1-2 & 7-10 \\
        OT Test                  & never & 1-2 & 7-10 \\
        Red-Teaming              & always & 3-4 & 30+ \\
        \bottomrule
        \end{tabular}
    }
    \end{table}
    
    Although different assignments have a similar project organization, their execution differs due to the respective client and target environment. Table~\ref{table_project_types} shows the main types of security assignments encountered during interviews.
    
    \textbf{Vulnerability Assessments} focus upon achieving a high coverage of the targeted assets, which are typically external IP-ranges (including web servers) or internal networks (including clients and internal infrastructure). Enumerating targets, e.g., through web crawling or network scans, leads to the creation of important inventory databases. Those are subsequently used to test against known vulnerability databases, known configuration errors or generic vulnerability classes such as SQL injections. As assignments typically include large amounts of potential targets, a high level of automation is necessary.

    \textbf{Penetration Tests} (Pen-Tests) share similarities with vulnerability assessments. The demarcation point between those two varied between interviewees. The situation is further complicated as vulnerability scans are often used as an initial step during pen-testing. Generally speaking, while vulnerability assessments focus on breadth, pen-testing focuses on depth, i.e., thoroughly breaking a single target. Pen-Tests are within the realm of application security: in addition to well-known vulnerabilities or configuration errors, new vulnerabilities are hunted within the software under test. Penetration tests are often performed against custom-written software where no prior vulnerabilities are published in vulnerability databases. As the scope is tight, customers commonly provide dedicated test environments against which destructive tests can be performed. Another benefit of the limited scope is that the execution of a penetration test can be highly structured, some ($n=2$) interviewees went as far as calling them ``\textit{catalog-based}''. Pen-tests are primarily performed manually. 
    
    \textbf{Internal Network Tests} verify the security and resilience of internal networks. Their basic assumption is ``\textit{assumed breach}'', i.e., the adversary is already within the local network and now attempts to gain sensitive data or achieve higher privileges --- emulating Ransomware scenarios that have recently scourged companies. Microsoft Active Directory (AD) is ubiquitous in corporate networks; thus, if present, it is the main target. In these cases, the security assignment's intent is to obtain domain administrator privileges. The focus lies on exploiting known vulnerabilities, product features, mis-configurations, and insufficient access-control or hardening measures. Another big aspect is Lateral Movement, i.e., using compromised systems to pivot to new targets. Assignments are made against productive environments.
    
    \textbf{OT Tests} target Operational Technology (OT) such as SCADA or ICS (Industrial Control System) networks. They can be differentiated into product tests and in-situ network tests of already configured systems. As solutions consist of off-the-shelf software that is highly customized for usage within the corresponding client network, the latter are often preferred by the customer. Tested subjects often use proprietary protocols; therefore, reverse engineering is a common practice in OT tests.
    
    OT facilities, e.g., power plants, are expensive and often hard to come by, thus a dedicated testing environment is rarely available. Testing commonly occurs during scheduled down-times; this severely impacts the available test window. Another related particularity: availability often trumps the breadth or depth of performed security tests. As test subjects are ``\textit{connected to the real world}'', negative side effects are potentially catastrophic. Security tests are therefore highly coordinated with customers to prevent any negative fallout. This often prohibits any covert action. Regulatory requirements~\cite{nisg} lead to a convergence between IoT and OT devices. In addition, Microsoft Active Directory starts to encreep OT networks, thus creating an overlap with Internal Network Tests.
    
    Compared to other approaches, in \textbf{Red-Teaming} the attackers have a concrete mission, e.g., gain access to a defined subset of computers or a source code repository. While during \textit{Internal Network Penetration Tests} gaining Domain Admin is often the final goal, this is only a means for achieving the mission during Red-Teaming. Attackers holistically target a company and employ additional techniques such as Open Source Intelligence (OSINT) and Social Engineering; Post-Exploitation is more prominent compared to other disciplines. Red teaming is not concerned with broad coverage, but with achieving the team's defined objective. Red-Teaming does not only attack the target's technical security posture but also the response of the blue team, i.e., defenders. Thus covert operations, hidden persistence, command\&control systems (C2) and evasion of defensive techniques enter the picture.
    
    Assignments are often performed in larger teams and over extensive time frames, making information transfer between participants more important. Adding additional team members to speed up an ongoing operation is problematic as the new team members do not share the existing member's target system knowledge.
    
    \subsection{Black- vs. Gray-Box Security Testing}
    \label{section:black_vs_gray}

    When it comes to test execution, an important distinction is the amount of information and support provided by the customer. During black-box tests, practitioners go in ``\textit{blind}''; no information except the scope is given. During white-box tests, full system access or even the source-code of the tested application is given. Gray-box tests lie in-between: often access credentials or system architecture descriptions are provided before testing commences.
    
    Pure white-box tests, as in ``source-code reviews'', are rarely performed due to their prohibitive costs. The type of assignment is also of importance: red-teaming is almost always performed as a black-box test as the target's personnel is not involved beneficially. OT tests are often performed in tight lock-step with customers (to reduce the potential fallout) and thus are gray-boxed. Interviewees overwhelmingly \textbf{recommended moving from black-box towards white-box testing}. The reasons given were time and thus cost efficiency, as well as potential for improved test coverage.
    
    In other areas, customers are helping pen testers to improve efficiency too. ``Assumed breach'' scenarios in Internal Network Penetration Testing conceptually assume that a client computer will be breached eventually and thus use a breached computer as a starting point for investigations. During web pen tests or during external scans, rate limits or firewalls are commonly disabled to allow swift pen test execution. During web application pen-tests, internal details, such as used technologies, are commonly provided to reduce the search space.
    
    \subsection{Typical Testing Workflows}
    \label{sub:workflow}
    
    Participants were asked to detail the execution of the different types of assignments. This section describes the peculiarities of the different areas.
    
    Activities performed during \textbf{Web Penetration Tests} can be separated into exploratory intuitive testing and exhaustive testing against checklists or standards. All interviewees utilized both, no specific ordering between those two was detected, although if the checklist-verification was automated, it often was run in parallel to exploratory testing. If a high-level of automation is achieved, the manual exploratory testing can be integrated into the automation: one interviewee detailed a multi-stage automated test-setup containing multiple enumeration steps, where the result of each step was manually verified, rectified and used to instrument subsequent automated steps. Manual testing, e.g., manual crawling, was integrated as an additional input into the tested steps.
    
    According to interviewees, most time and effort are spent upon authorization tests. An application typically has multiple user groups with different access rights. During testing, penetration testers request one or more users per existing group and try to perform unauthorized data access with one user using data of another user. To verify responses, testers need documentation about the implemented access groups. If none was given, interviewees approximate a model of the access rules through probing/testing and experience.
    
    With the exception of testing for authentication or authorization, automated testing was deemed well-established and automated tooling was commonly employed. Common injection attack vectors were well covered by tooling, for example, \textit{sqlmap}~\cite{sqlmap} for testing for SQL injections. Multiple Web-Application-Testers ``complained'' that typical injection-based attacks which were common 10 years ago are now seldom seen and are rather used for illustrative purposes during education. Their suspected ``culprit'' is the rise of web application frameworks with sane defaults that automatically prevent many attack classes. Multiple interviewees considered switching their area of interest due to this development.
    
    Multiple interviewees described API-based tests as tedious. Typically an API test is performed by calling a sequence of operations. Each operation is detailed through an API specification provided by the customer, e.g., through OpenAPI/Swagger or WSDL files. In theory, directly testing the back-end API reduces the pen-testing overhead as the tester can focus upon the core functionality; in practice, API tests become time-consuming due to a lack of documentation with sufficient quality. API documentation only describes single operations, often lacking detailed descriptions of valid input formats and their semantics. In addition, to achieve good test coverage, test cases need to perform a sequence of causally dependent API calls, potentially reusing and refining data between operations. While performing a traditional web application pen test, this causality and examples of input data can be derived from the captured web traffic. When performing API tests, these have to be derived from the API specifications or, more realistically, by pestering the customer's liaison contact.
    
    \textbf{Internal Network Tests} often occur in phases which are ordered from ``\textit{quiet}'' to ``\textit{loud}'' when it comes to visibility. A typical assignment targeting a Microsoft Active Directory might include the following phases: initially, only network access is granted. The attacker either sniffs the network for exposed access credentials or utilizes MitM- and spoofing attacks to gain user credentials or tokens. In addition, anonymously accessible network shares are investigated for ``\textit{juicy}'' information such as user or admin credentials. Exploits are used against vulnerable network services if the risks of detection and stability are deemed acceptable. In the second phase, an attacker has either already gathered user credentials or has been provided with those by the customer. These credentials are typically for non-privileged domain users, and attackers utilize them to further enumerate shares, gain access to additional domain accounts or computers, or gain local administrative privileges. Lateral Movement often incurs during this phase. In the next phase, the attacker has either gained or is provided local administrative privileges and tries to perform further Lateral Movement until a domain administrative account is compromised. With that, the whole network is owned.
    
    Please note, that phases do not follow a traditional waterfall model. According to interviewees ($n=2$), often the domain admin credentials can be gathered during the initial phase. This is then noted, and additional attacks are performed until the agreed upon timebox is reached.
    
    Many automated attacks, e.g., EternalBlue~\cite{boyanov2018educational} or certify~\cite{certify}, were described as ``\textit{too loud}'' or ``\textit{unstable}'' for use during the initial phases. Another automation topic was the identification of ``\textit{juicy}'' files within network shares: this activity is performed primarily manually as the identified data are context specific. In addition, creating a full-copy of a network share is time- and network-sensitive as well as easily detectable, and countermeasure systems using honey-tokens are beginning to be deployed at customers' sites.
    
    All interviewees in the IoT area mentioned applying industrial standards as well as the usage of checklists that included the OWASP IoT
~\cite{owasp_isvs} and OWASP Firmware Testing guides~\cite{owasp_firmware}.
    
    \textbf{Red-Teaming} is special due to its evasion- and deception-based methods as well as through its objective-based approach. A red team initially has knowledge of its objective, e.g., gain access to a special server in department X, as well as a broad allowed scope, e.g., the targeted company. Teams initially model how to breach the company, e.g., by identifying potential social engineering victims. After the breach, low-key enumeration is used to covertly model ``\textit{how a company works}'' and then abuse that knowledge to derive attacks that mirror expected traffic and behavior patterns. Throughout a red-teaming campaign, a map of known or breached elements is built and compared to the imagined map of the company that includes the final objective: if both converge, the objective should be achieved.
    
    Automation employed for network lateral movement or breaching web applications originate from the other pen-testing disciplines but have to be re-evaluated against their chance of being detected. As red-team assignments are performed against real and live systems, the scope of destructive operations might be limited.
    
    \textbf{OT-Tests} have their own challenges. Due to the prevalence of proprietary protocols, time-consuming reverse engineering of those protocols often occurs. Mentioned experiences of our interviewees indicate that Security-by-Obscurity is still common; this would match the perceived resistance of some ICS suppliers when faced with responsible disclosure requests. Due to the time burden of reverse-engineering, it frequently has to be aborted due to the timeboxed nature of testing.
    
    Due to the potentially catastrophic side-effects of testing, a risk-based approach is often applied: together with the customer, a threat model workshop can be performed, and potential scenarios that warrant testing identified. Those scenarios, and only those, are subsequently manually executed against the OT system. As the available amount of time is fixed, threat modeling and performing the derived tests compete for the same temporal resources.
    
    \subsection{Automation}

\begin{table}[!t]
    \caption{Commonly Named Tools.}
    \label{table_common_tools}
\begin{minipage}{\columnwidth}
\begin{center}

    \resizebox{\columnwidth}{!}{
    \begin{tabular}{lllr}
    \toprule
    Tool & Area & Availability & \# \\
    \midrule
    PortSwigger BURP Suite~\cite{burp} & Web-Testing & free, commercial & 7 \\
    BloodHound~\cite{bloodhound} & AD Enumeration & OSS & 5 \\
    SQLMap.py~\cite{sqlmap}  & Web/SQLi & OSS & 3 \\
    nmap~\cite{nmap}       & Network & OSS & 7 \\
    nessus~\cite{nessus}     & Network & commercial & 8 \\
    gobuster~\cite{gobuster}, dirbuster~\cite{dirbuster} & Network & OSS & 2 \\
    certify~\cite{certify}    & AD Exploitation & OSS & 4 \\
    metasploit~\cite{metasploit} & Exploitation & OSS & 3 \\
    nuclei~\cite{nuclei} & Exploitation & OSS & 3 \\
    \bottomrule
    \end{tabular}
 }

\end{center}
 \bigskip
 \begin{center}
 \footnotesize \# denotes the interviewee count mentioning the corresponding tool.
 \end{center}
\end{minipage}
\end{table}
    
\begin{table*}[!t]
    \caption{Excerpt of sub-themes of ``Identifying Vulnerable Areas or Operations''}
    \label{table_identifying}

\begin{minipage}{\textwidth}
\begin{center}

    \begin{tabular}{lrp{13cm}}
    \toprule
    Subtheme & \# & Representative Quotes \\
    \midrule
    High-Level Targeting & 7 & \textit{``We select the attack that would be the most cost-effective for the attacker''}\\
    & & \textit{``\ldots before we attack proprietary protocols we'll attack a windows domain server missing updates.''}\\
    Experience & 12 & \textit{``How we actually work? We look for obvious vulnerabilities, those that jump out immediately''}\\
    & & \textit{``I know that from my time programming C/C++\ldots I find the errors that I made back then''}\\
    & & \textit{``I search for vulnerabilities that I have seen and exploited before.''}\\
    & & \textit{``\ldots often I see systems that I have already seen when doing CTFs\ldots then I already know how to attack it''}\\
    Familiarity with Target & 4 & \textit{''If it is a repeat customer then you already know how they tick and what their problems are''}\\
    Observed Features & 10 & \textit{``One runs through the web applications and sees a feature and thinks ``this looks interesting, could it be implemented weirdly''?''}\\
    & & \textit{``If there's an upload function, I am interested.''}\\
    Observed Technology & 11 & \textit{``Some things cannot be done securely, for example PHP.''}\\
    & & \textit{``Well, you always feel happy when the application is somehow a PHP application.''}\\
    Modeling Behavior & 9 & \textit{``Testing is manual, as you need to get a feel how the application is supposed to work and answer''}\\
    & & \textit{``You search for unexpected behavior\ldots for example a database that throws an error when you enter a '. ''}\\
    Intuition & 8 & \textit{``This will be esoteric\ldots but I believe there is some organ that tingles if an operation looks fishy''}\\
    \bottomrule
    \end{tabular}
\end{center}
\bigskip
\begin{center}
\footnotesize Subthemes mentioned by interviewees, \# denotes the interviewee count.
\end{center}
\end{minipage}
\end{table*}

    All interviewees used pre-made tooling, while few ($n=3$) wrote additional tooling on their own. Overall, the tooling situation for specific testing areas was seen in a positive light. In contrast, ``\textit{all-in-one}'' tools were seen in a negative light. Multiple interviewees remarked that a ``\textit{fully automated tool cannot replace a pen-tester}'' or, as one interviewee cynically replied, ``\textit{yeah, I want a tool where I can click a button and magically I get a finished pen-test report}''. Practitioners relied on multiple small tools for different areas, e.g., \textit{gobuster}~\cite{gobuster} for content discovery or \textit{sqlmap}~\cite{sqlmap} for testing SQL injections. PortSwigger's \textit{BURP Proxy Suite}~\cite{burp} was used by every web application pen-tester interviewed. See Table~\ref{table_common_tools} for a list of commonly named automated tools.
    
    \textbf{Problems with tooling.} Interviewees remarked that the setup overhead of automation tools can be problematic. Especially for short-term projects,  such as vulnerability assessments or tightly-timed web application pen-tests, the initial setup overhead and processing time can be prohibitive for deploying tooling. Another problem was coverage: even within the same problem area, the coverage of different tools widely diverges, and the situation is made worse as commonly no tool provides full coverage of a testing area. To counteract this, practitioners commonly use multiple tools redundantly, yielding more processing time overhead and needing manual merging of the different tools' results.
    
    Some areas were described as not suitable for automation. As OT systems are finicky and the potential fallout catastrophic, automated tests are often not feasible. Additionally, when performing social engineering during red-team assignments, fully automated tools are avoided for both fear of detection and ethical qualms because they would be used on human targets.
    
    \textbf{Extendability and Community} was identified as an important discriminator by practitioners. Both are related to fast-paced developments within the exploit community: if a tool can be proactively extended or be scripted by the community, it and its implemented methods can evolve faster compared to reactive development within walled gardens. An example of an OSS tool utilizing community-provided detection rules is \textit{nuclei}~\cite{nuclei}; an example of a commercial tool with good OSS extendability is the PortSwigger \textit{BURP Proxy Suite}~\cite{burp} with its integrated \textit{BApp Store}.
    
    \textbf{Manual fine-tuning to reduce search space.} Multiple interviewees mentioned that they are adjusting the tooling according to their ongoing findings. Examples of this feedback loop would be limiting tested vulnerability classes to feasible ones, e.g., not testing a static website for SQL injections, or limiting tested database queries to concrete database dialects.
    
    \section{How do Hackers think?}
    \label{section:mind}
    
    While Section~\ref{pentests} describes the external view on pen-tests, their type and activities activities performed during them, this section focuses on the inner workings and thoughts of security professionals during testing, detailing their decision processes and potential sources of their intrinsic motivation.
    
    \subsection{Exploiting Configuration vs. Applications}
    \label{section:conf_vs_code}
    
    A reoccurring theme was the distinction between searching for known vulnerabilities and hunting for new vulnerabilities.
    
    Examples of the former would be executing a known vulnerability scan against off-the-shelf software or investigating a Microsoft Active Directory for misconfigurations; an example of the latter would be searching for SQL injections within a custom written application or discovering a new vulnerability class.
    
    Synonyms given for ``\textit{searching for known vulnerabilities vs. hunting for new vulnerabilities}'' were ``\textit{vulnerability assessments vs. application security}'' or ``\textit{hacking configuration vs. hacking programs}''.
    
    These two categories are fluid. For example, findings from ``\textit{hunting for bugs}'', i.e., a new 0-day exploit against a software, can end up within ``\textit{searching for known vulnerabilities}'', i.e., when a rule for detecting 0-day is added to a web vulnerability scanner.
    
    While not stated explicitly during the interviews, we assume that our interviewee's mental model is primed through their understanding of this divide, and highly impacts tool and technique selection. As an interviewee mentioned, ``\textit{you don't hunt for 0-days during an Active Directory assignment}''. This implies that pen-testers will not consider spending days fuzzing a domain controller for new vulnerabilities during internal network scans.
    
    \subsection{Identifying Vulnerable Areas or Operations}

    Participants often described exploratory testing during which they were guided by intuition. Through follow-up questions, further information about this intuition was gathered.
    
    All interviewees were analyzing requests and responses; the former for conspicuous parameters and the latter for occurrences of error messages or other suspicious behavior, that is, behavior that does not fulfill the testers' expectations.

    During the interviews, multiple areas were identified where security testers possessed a mental model of the expected behavior of the software-under-test; during testing security testers were trying to find operations that could trigger unexpected behavior which, in turn, might turn into a security vulnerability. Those mental models were built from experience, e.g., prior assignments or experience within the specific business area, as well as adapted during the security test itself, e.g., \textit{``learning how the application works''}. A summary of multiple observed mental models is shown in Table~\ref{table_models}.
    
    \begin{table*}[t]
    \caption{Excerpt of observed models}
    \label{table_models}
    \centering
    \begin{tabular}{lllll}
    \toprule
    Area & Input & Identified Elements & Describes & Used for \\ 
    \midrule
    Web Testing & Web Traffic & Access Rules & ACL model & Authentication Checks \\
    Red-Teaming & Network Traffic & Communication Patterns & Expected Communication & Covert Channels \\
    Network Tests & local data and network shares & File data and metadata & Company Data & find juicy information \\
    OT tests & data flows & data flow model & system architecture & identify test scenarios \\
    OT tests & network traffic & network commands & network protocol & protocol reversing \\
    Web Testing & web traffic, context & used technologies & technology stack &  potential vulnerabilities \\
    Web Testing & web traffic & HTTP requests and responses & input model & generate tests \\
    \bottomrule
    \end{tabular}
    \end{table*}
    
    Pen-testers attributed their intuition to experience which could be built from previous penetration tests, participation in CTF events, prior engagements with the same client or industry area, or by implementing similar software solutions during their former life as software developers. Participants remarked that during testing, they are triggered by vulnerabilities or exploits they had recently read about and, in response, would start additional research. One penetration tester mentioned creating a topic map during everyday research which they then refer back to during assignments.
    
    Related to experience, practitioners had preconceptions about the technologies used or features implemented. Some functionality, e.g., file uploads or XML processing, were thought to be hard to implement in a secure manner --- to quote a participant, ``\textit{there are some things that just cannot be implemented correctly}''. Similar resentments were discovered about used technologies. Some programming languages were deemed to increase the probability of an application containing defects; an interviewee mentioned thinking ``\textit{let's see how developers have been fooled again}'' when going into assignments. As cynical as it may be, PHP was often mentioned as such a technology.
    
    Two distinct positions were experienced regarding the learnability of this intuition. On one side, ``\textit{nobody is born a super hacker}'', on the other hand, one interviewee mentioned that the best penetration testers in their peer group exhibited hacking-style behavior already during kindergarten. Debating nature-vs-nurture or art-vs-craft would go beyond the scope of this publication. Regardless of this, common consensus was found that hacking skills are improved through practice.
    
    It is important to note that participants may be subject to \textit{selection} and \textit{survivorship} bias. They might find vulnerabilities in areas they focus on, ignoring plentiful vulnerabilities in other areas they are historically ignoring. After a vulnerability has been found in an area, the increased attention upon that area often yields multiple subsequent vulnerabilities~\cite{more_printnightmare}.
    
    \subsection{Dealing with Uncertainty}
    
    Pen-testers routinely have to deal with uncertainty as they lack transparency of the tested system: pen-testers must make assumptions about requirements, the tested system's architecture, as well as about accepted input values and the corresponding expected output parameters~\cite{trubiani2019performance}. They evaluate those against their expectations, and if a system deviates, examine the deviation for exploitability. When in doubt, testers can escalate and query their clients, but this is deemed to be time-inefficient and thus minimized.
    
    Examples of uncertainty would be a pen-tester issuing a HTTP request where they expect an ``access denied'' response but instead, receive a successful response containing data that cannot be clearly classified as belonging to the current user or not. Another example would be testing for time-based blind SQL injection vulnerabilities where the measured latency is not sufficiently deterministic for verifying the vulnerability. Similarly, second-order attacks cannot easily be attributed to the initial request but only to the operation that eventually contained the vulnerability.
    
    Penetration testers modify existing valid requests to include malicious payloads. When these requests produce errors, the reason can be uncertain: was it a potential vulnerability? A successful input filtering algorithm? Or an application error that cannot be exploited? This classification impacts the selection of subsequent requests and attacks.
    
    Another instance of uncertainty occurs during tool optimization: tool output is continuously used to further optimize subsequent tool invocations. Interviewees performed a sanity check if reported system fingerprints were feasible and forfeited them otherwise. In addition, some high-impact decisions, such as limiting the expectations to a single DBMS type, were verified with the client before incorporating them into tooling selection or configuration.
    
    \subsection{Don't waste my time}
    
    One theme discovered was that interviewees feel the need to be time-efficient. This might be related to tight time-budgets or very constrained test-bed availability being anathema to good test coverage. Shortcuts were taken to reduce menial tasks. For example, during internal network tests, a breach is already assumed. The interviewees defended this decision through ``\textit{this will eventually happen through social engineering anyways}''. A similar argument was given for being provided accounts with local administrative privileges: ``\textit{a real attacker can just wait for the next 0-day}'', or for disabling Anti-Virus solutions as evading them ``\textit{takes time not skill}''. Tests with foregone conclusions were considered tedious, one example given was testing an Anti-Virus solution embedded within a web-application with different payloads. The repetitiveness of this task might contribute to this too. This aversion to responsible disclosure procedures might be correlated to bad experiences during prior disclosures: the vendor's responses were mostly ``wasting'' the interviewee's time.
    
    \subsection{Quality Control}
    
    Pen-Testers were concerned about the quality of their work, especially when working with high-stakes data such as health records --- ``\textit{nobody wants to be that pen-tester that overlooked a vulnerability that was later exploited}''. A tester's attention is also a limited resource: at least one pen-tester remarked that web application tests can be monotonous and that after 3--4 days their motivation degrades. Usage of checklists, automated baseline scans, and working in teams were encountered as quality improvement measures.
    
    The applicability of checklists depends upon the testing domain. Some domains, e.g., Web-Applications or Mobile Applications, were seen as narrow and thus supporting the creation of security checklists. Other domains such as IoT were described as diverse and impeding the creation of a unified security checklist.
    
    Checklists were often derived from open industry standards; they were maintained and extended by companies, but the resulting in-house checklists were seldom given back to the community and published. Common base for checklists was the OWASP trifecta of Vulnerability Top 10, Software Verification Standard and Testing Guide; instances of those are provided by OWASP for multiple domains such as Web-Applications~\cite{owasp_top10,owasp_otg,owasp_asvs}, Mobile Applications~\cite{owasp_mtg, owasp_masvs}, IoT~\cite{owasp_isvs} or Firmware~\cite{owasp_firmware}. Surprisingly, neither MITRE ATT\&CK\textsuperscript{\textcopyright}~\cite{strom2018mitre} nor PTES~\cite{ptes} were mentioned by our interviewees.
    Working in teams or asking colleagues can be seen as a broadening of the available experience pool or as employing a ``human checklist''.  The use of automated tools as baseline scans that upheld minimal quality standards can also be interpreted as quality control. Interviewees mentioned usage of fully-automated commercial web vulnerability scanners such as NetSparker~\cite{NetSparker} or Acunetix~\cite{Acunetix} for this purpose. Some HTTP interception proxies, for example, PortSwigger BURP~\cite{burp} or OWASP ZAP~\cite{zap}, have gained similar scanning capabilities. Those were used by some of the interviewees and encroached on terrain traditionally taken by web vulnerability scanners.    
    In defense of testers, full coverage of the software-under-test is not feasible due to the black- to gray-box nature of security assignments.

    \begin{table*}[!t]
    \caption{Excerpts of \textit{Dealing with Change}: How is Security-Testing changing?}
    \label{Future_Directions}

\begin{minipage}{\textwidth}
\begin{center}
    
    \begin{tabular}{p{3cm}rp{13.5cm}}
    \toprule
    Sub-Theme & \# & Representative Quotes \\
    \midrule
    Impact of Frameworks& 5 & \textit{``Security improves because frameworks help developers write secure code''}\\
    & & \textit{``Pen-Testing has become boring as critical vulnerabilities are found less often''}\\
    & & \textit{``Usage of secure frameworks pushed vulnerability hunting towards business logic.''}\\
    Defensive Mindset& 3 & \textit{``Developer awareness about security has become better.''}\\
    Changing targets & 7 & \textit{``In the future we might use social engineering not only for the initial foothold, but also for lateral movement''}\\
    & & \textit{``Rich-client applications are still fun\ldots they feel like web applications twenty years ago.''}\\
    & & \textit{``Active-Directory: I moved into this area because it is fun to break into a system within days.''}\\
    & & \textit{``The situation in OT will stay the same. It's hard to modernize all the legacy hard- and software.''}\\
    & & \textit{``Some OT networks are ransomware-ready.''}\\
    \bottomrule
    \end{tabular}
    
\end{center}
\end{minipage}
    
    \end{table*}

\subsection{Dealing with Change}

    Security is in a constant state of flux. Compared to other disciplines, the existence of active adversaries --- the struggle between red and blue teams --- lead to a Red Queen's Race: participants must run to stand still~\cite{harang2018measuring,bukac2014red}. If not evolving, the respective adversary will overcome.
    
    Interviewees lamented that some areas --- breaking into web applications, breaching external infrastructure/perimeters, and reverse-engineering --- have become harder due to boosted defenses such as usage of frameworks, improved default configurations, and heightened awareness of security posture (cf. Table~\ref{Future_Directions}). They are partially switching work areas, i.e., turning towards OT or internal network testing.

    \section{Discussion and Implications}

We review our findings following the structure of our initial research questions to formulate points of discussion and implications for security researchers and practitioners.

\subsection{Alignment between Research and Industry}

We started this study with two questions, \textbf{``What do common security tests look like?''} and \textbf{``How do Hackers perform their tasks?''}. Those questions were broadly formulated to gain insight into how common assignments for practitioners look like, and how practitioners navigate their tasks within those assignments. These questions were particularly motivated by the fact that existing work is not grounded in the realities of offensive security practices.

\subsubsection{Research must match a Project's Scope.} During interviews, we identified typical security assignments with their respective typical resource allocations. Research should heed those resources allocated. For example, when targeting web vulnerability assessments, a typical project was given with 2--4 days of manual effort. Setting up a fuzzing pipeline, running the fuzzer, and analyzing its results is not feasible in this short time frame, thus rendering generic fuzzing rather infeasible for web security practitioners. Still, searching Google Scholar for ``\textit{fuzzing web applications}'' yields 23000 results.

Given that interviewees mentioned the prevalence of web application frameworks and their preference for grey-box testing, SBOM-based solutions should be a better fit and would warrant additional research.

Most assignment types were done in solitary or as a paired team, indicating that research into collaborative solutions might be of limited use. The one exception using larger teams was Red-Teaming although here collaborative solutions integrated into C2-frameworks are already commonly used.

Automation with direct target-interactions were deemed problematic in the Red-Teaming and OT areas due to the sensitivity of their targets. In OT, security by obscurity still seems to be common, limiting the opportunities for source-code analysis based approaches. On the other hand, improvements to reverse-engineering binaries or protocols would be appreciated by practitioners.

Recently, the usage of Large Language Models (LLMs) for automated security testing has been explored~\cite{getting_pwned}. While preliminary results look promising, to maximize their long-term impact the resulting automations should be aligned to the mentioned industry issues.

\subsubsection{Security Researchers and Security Practitioners.} Separating security into academic research and industry creates a false dichotomy. Industry itself is, at least, separated into security practitioners and security researchers. The former are practitioners that perform customer-specific assignments: those are the people that typically perform short-term penetration tests and directly communicate with clients to improve their security. In contrast, security researchers do not exclusively work on short-term client projects but spent time researching new attack techniques and vectors. An example of the former would be an anonymous pen-tester working on a different web-application every week; an example of the latter would be James Kettle investigating and documenting a new attack class, HTTP Request Smuggling, over many years~\cite{http_smuggling, http_smuggling_2}. Security researchers search for new attack vectors or analyze a software product for a prolonged period of time to release exploits or be awarded CVEs. Security Practitioners are more focused on hunting configuration errors, exploiting well-known vulnerabilities, or identifying new instances of known attack classes. They utilize information and tools from security researchers for that.

Tools such as fuzzers are thus more applicable to security researchers than to security practitioners. The large amount of research into fuzzing indicates that academic research is targeting security researchers rather than practitioners and thus are only indirectly improving the security landscape when information from security researchers trickles down to practitioners.

\subsection{Opportunities for Research.}

We now want to answer the important final question, \textbf{``What tedious or time-consuming areas could be improved?''} throughout the rest of this section and frame them as opportunities for future research that directly benefits security practitioners.

\subsubsection{\textbf{Automating Authorization Testing}}
For security tests with a relatively restricted scope such as \textit{web application tests}, we suggest research into covering additional vulnerability classes. \textbf{Authorization Testing} is currently performed manually and was named one of the most time-consuming parts of testing and thus would be a fruitful target for automation research. Current gaps are manifold: detection of potential operations, accepted parameters, and potentially malicious parameters; generation of payloads as well as the assessment of an attack's success. A subtle problem is the classification of returned web pages and downloads into authorized and unauthorized content as this is highly context specific.

\subsubsection{\textbf{Gray-box Testing}}
The preference for gray-box testing by software security professionals was surprising and can have a significant impact on software testing design: if the target's configuration or source code can be accessed (or if the target is willing to instrumentalize the target software through sensors as is done in IAST), \textbf{automated software testing approaches using source-code or configuration} become increasingly feasible for security testing. Further research into automated source code and configuration file analysis from a security perspective, is currently underexplored and ripe for investigation. Research in this area yields dual-use tools, aiding both offensive security professionals searching for vulnerabilities as well as defensive software developers trying to prevent vulnerabilities from entering their code in the first place.

\subsubsection{\textbf{API Workflow Discovery for Security Test Generation}}
Interviewees lamented that the manual creation of API security test-cases is a tedious and time-consuming process. While the automation of API test generation would be advantageous, the following gaps currently prevent this: discovery of API endpoints and operations, generation of benign requests as a baseline, combining single requests into test flows using social and semantic information, deriving malicious test cases, and finally evaluating test outcomes. The \textbf{automatic generation of security test suites based upon API definitions and traffic patterns} would reduce testers' odium for utilizing this important class of testing. While there have been several works that propose approaches for API discovery~\cite{yessenov2017demomatch, torres2011improving}, the kind of discovery we envision would focus on maximizing coverage for security tests.

\subsubsection{\textbf{Information Discovery for Security Testing}}
\textit{Internal Network Tests} and \textit{Red-Teaming} are highly dependent on discovering and utilizing client-specific information. \textbf{Stealthy information gathering from compromised systems or network shares} is performed manually, and thus its efficiency could be improved. The goal is the automated identification of ``juicy'' information while reducing the number of read requests to minimize network impact or the chance of triggering intrusion detection systems. Research in this area would also benefit defenders as it would make forensic work, e.g., analyzing data breaches, more efficient.

\subsubsection{\textbf{Scaling Personalized Phishing with ML}}
\textit{Phishing} is an important part of the red-teaming workflow and is commonly done manually, due to the nature of customization proper phishing requires. We see an opportunity to investigate the \textbf{increase of scalability of social engineering through machine learning} techniques. To create highly effective phishing mails, currently, mails are manually customized to fit the respective recipient. Machine learning techniques could automate this and thus provide \textbf{Spear Phishing at Scale}, as they have already been shown to personalize natural language communication in other domains~\cite{ferretti2016automatic, katakis2009adaptive, xu2018deeptype}. An additional avenue for research is the identification of potential targets for social engineering, both from an external perspective (identifying initial recipients within a company) as well as \textbf{detecting informal networks within companies to enrich subsequent social-engineering campaigns} --- this is an example of the red-teaming theme of ``\textit{understanding how companies function}''.

\subsubsection{\textbf{Human-in-the-loop for OT testing}}

OT professionals were weary of fully automated security tests due to the potential negative impact on stability and thus availability. We suggest research into supplemental areas while letting humans decide which attacks to execute. One example would be to \textbf{reduce the pain and effort of reverse engineering protocols}: OT tests are very time-bound thus there is little time for fuzzing or reverse-engineering OT protocols while the potential benefit might be immense due to security being provided by the obscurity of those protocols. Combining fuzzing with automatic reverse-engineering should yield large benefits~\cite{gascon2015pulsar}.
The fear of potential fall-out has other consequences too: OT-tests are often performed by executing scenarios in lockstep with the customer. The scenarios are identified through threat modeling components and their data flows. To reduce the time spent on this effort, ways of \textbf{automatically deriving scenarios including attack paths} from system and data flow diagrams should be investigated.

Both OT professionals and red-teams were weary of fully automated testing solutions due to the potential negative impact upon stealth (red-teaming) or stability (OT). To facilitate the deployment of automated systems, \textbf{research into Human-Computer Interactions to bolster the acceptance of ML and automated systems} is needed. It is assumed that important topics will include humans-in-the-loop as well as the explainability of automated reasoning.

\subsubsection{\textbf{Studying Knowledge Communities for Security Testers}}
Our interview participants unsurprisingly felt the need for ongoing education w.r.t. new vulnerabilities and security trends. They synthesized information from multiple sources, the pivotal one being Twitter/X. Research on how developers stay current~\cite{singer2014software} and how development communities shape around news outlets~\cite{aniche2018modern} should be extended to the security arena, especially now that recent stewardship changes at Twitter might impact its reach. \textbf{Automated recommender systems utilizing diverse hacking news sources} such as news outlets, social media, and, the ``darknet'' should enable security professionals to stay up to date easier.

\section*{Acknowledgment}

We thank the anonymous interview participants for their time, and Loren Kohnfelder and Geraldine Fitzpatrick for providing feedback.

\balance

\bibliographystyle{ACM-Reference-Format}
\bibliography{thebibliography}


\begin{thebibliography}{65}


\ifx \showCODEN    \undefined \def \showCODEN     #1{\unskip}     \fi
\ifx \showDOI      \undefined \def \showDOI       #1{#1}\fi
\ifx \showISBNx    \undefined \def \showISBNx     #1{\unskip}     \fi
\ifx \showISBNxiii \undefined \def \showISBNxiii  #1{\unskip}     \fi
\ifx \showISSN     \undefined \def \showISSN      #1{\unskip}     \fi
\ifx \showLCCN     \undefined \def \showLCCN      #1{\unskip}     \fi
\ifx \shownote     \undefined \def \shownote      #1{#1}          \fi
\ifx \showarticletitle \undefined \def \showarticletitle #1{#1}   \fi
\ifx \showURL      \undefined \def \showURL       {\relax}        \fi
\providecommand\bibfield[2]{#2}
\providecommand\bibinfo[2]{#2}
\providecommand\natexlab[1]{#1}
\providecommand\showeprint[2][]{arXiv:#2}

\bibitem[Acu({[n.\,d.]})]%
        {Acunetix}
 \bibinfo{year}{[n.\,d.]}\natexlab{}.
\newblock \bibinfo{title}{Acunetix: Web Vulnerability Scanner}.
\newblock \bibinfo{howpublished}{\url{https://www.acunetix.com/}}.
\newblock
\newblock
\shownote{Accessed: 2022-09-30}.


\bibitem[blo({[n.\,d.]})]%
        {bloodhound}
 \bibinfo{year}{[n.\,d.]}\natexlab{}.
\newblock \bibinfo{title}{BloodHoundAD: Six Degrees of Domain Admin}.
\newblock
  \bibinfo{howpublished}{\url{https://github.com/BloodHoundAD/BloodHound}}.
\newblock
\newblock
\shownote{Accessed: 2022-09-30}.


\bibitem[ire({[n.\,d.]})]%
        {ireland_conti}
 \bibinfo{year}{[n.\,d.]}\natexlab{}.
\newblock \bibinfo{title}{{Conti cyber attack on the HSE}, Independent Post
  Incident Review}.
\newblock
  \bibinfo{howpublished}{\url{https://www.hse.ie/eng/services/publications/conti-cyber-attack-on-the-hse-full-report.pdf}}.
\newblock
\newblock
\shownote{Accessed: 2022-09-30}.


\bibitem[con({[n.\,d.]})]%
        {conti}
 \bibinfo{year}{[n.\,d.]}\natexlab{}.
\newblock \bibinfo{title}{Conti’s Hacker Manuals — Read, Reviewed \&
  Analyzed}.
\newblock
  \bibinfo{howpublished}{\url{https://www.akamai.com/blog/security/conti-hacker-manual-reviewed}}.
\newblock
\newblock
\shownote{Accessed: 2022-09-30}.


\bibitem[del({[n.\,d.]})]%
        {delve}
 \bibinfo{year}{[n.\,d.]}\natexlab{}.
\newblock \bibinfo{title}{Delve: Software Tool to Analyze Qualitative Data}.
\newblock \bibinfo{howpublished}{\url{https://delvetool.com/}}.
\newblock
\newblock
\shownote{Accessed: 2022-10-01}.


\bibitem[dir({[n.\,d.]})]%
        {dirbuster}
 \bibinfo{year}{[n.\,d.]}\natexlab{}.
\newblock \bibinfo{title}{DirBuster}.
\newblock \bibinfo{howpublished}{\url{https://www.kali.org/tools/dirbuster/}}.
\newblock
\newblock
\shownote{Accessed: 2022-09-30}.


\bibitem[cer({[n.\,d.]})]%
        {certify}
 \bibinfo{year}{[n.\,d.]}\natexlab{}.
\newblock \bibinfo{title}{GhostPack/Certify: Active Directory certificate
  abuse.}
\newblock \bibinfo{howpublished}{\url{https://github.com/GhostPack/Certify}}.
\newblock
\newblock
\shownote{Accessed: 2022-09-30}.


\bibitem[gob({[n.\,d.]})]%
        {gobuster}
 \bibinfo{year}{[n.\,d.]}\natexlab{}.
\newblock \bibinfo{title}{gobuster: Directory/File, DNS and VHost busting tool
  written in Go}.
\newblock \bibinfo{howpublished}{\url{https://github.com/OJ/gobuster}}.
\newblock
\newblock
\shownote{Accessed: 2022-09-30}.


\bibitem[mor({[n.\,d.]})]%
        {more_printnightmare}
 \bibinfo{year}{[n.\,d.]}\natexlab{}.
\newblock
  \bibinfo{title}{https://nakedsecurity.sophos.com/2021/07/16/more-printnightmare-we-told-you-not-to-turn-the-print-spooler-back-on/}.
\newblock
  \bibinfo{howpublished}{\url{https://nakedsecurity.sophos.com/2021/07/16/more-printnightmare-we-told-you-not-to-turn-the-print-spooler-back-on/}}.
\newblock
\newblock
\shownote{Accessed: 2022-10-03}.


\bibitem[Net({[n.\,d.]})]%
        {NetSparker}
 \bibinfo{year}{[n.\,d.]}\natexlab{}.
\newblock \bibinfo{title}{Invicti: Web Application Security for Enterprise}.
\newblock \bibinfo{howpublished}{\url{https://www.invicti.com/}}.
\newblock
\newblock
\shownote{Accessed: 2022-09-30}.


\bibitem[met({[n.\,d.]})]%
        {metasploit}
 \bibinfo{year}{[n.\,d.]}\natexlab{}.
\newblock \bibinfo{title}{Metasploit: Penetration Testing Software}.
\newblock
  \bibinfo{howpublished}{\url{https://github.com/rapid7/metasploit-framework}}.
\newblock
\newblock
\shownote{Accessed: 2022-09-30}.


\bibitem[bur({[n.\,d.]})]%
        {burp}
 \bibinfo{year}{[n.\,d.]}\natexlab{}.
\newblock \bibinfo{title}{Methodology for Top 10}.
\newblock
  \bibinfo{howpublished}{\url{https://groups.google.com/a/owasp.org/g/leaders/c/pFLxDLE28ZA}}.
\newblock
\newblock
\shownote{Accessed: 2022-09-30}.


\bibitem[nes({[n.\,d.]})]%
        {nessus}
 \bibinfo{year}{[n.\,d.]}\natexlab{}.
\newblock \bibinfo{title}{Nessus Vulnerability Assessment Solution}.
\newblock
  \bibinfo{howpublished}{\url{https://www.tenable.com/products/nessus/nessus-professional}}.
\newblock
\newblock
\shownote{Accessed: 2022-09-30}.


\bibitem[nma({[n.\,d.]})]%
        {nmap}
 \bibinfo{year}{[n.\,d.]}\natexlab{}.
\newblock \bibinfo{title}{Nmap: the Network Mapper --- Free Security Scanner}.
\newblock \bibinfo{howpublished}{\url{https://nmap.org}}.
\newblock
\newblock
\shownote{Accessed: 2022-09-30}.


\bibitem[nuc({[n.\,d.]})]%
        {nuclei}
 \bibinfo{year}{[n.\,d.]}\natexlab{}.
\newblock \bibinfo{title}{Nuclei: Fast and customizable vulnerability scanner
  based on simple YAML based DSL.}
\newblock
  \bibinfo{howpublished}{\url{https://github.com/projectdiscovery/nuclei}}.
\newblock
\newblock
\shownote{Accessed: 2022-09-30}.


\bibitem[zap({[n.\,d.]})]%
        {zap}
 \bibinfo{year}{[n.\,d.]}\natexlab{}.
\newblock \bibinfo{title}{OWASP Zed Attack Proxy (ZAP)}.
\newblock \bibinfo{howpublished}{\url{https://www.zapproxy.org/}}.
\newblock
\newblock
\shownote{Accessed: 2022-09-30}.


\bibitem[pte({[n.\,d.]})]%
        {ptes}
 \bibinfo{year}{[n.\,d.]}\natexlab{}.
\newblock \bibinfo{title}{PTES Technical Guidelines}.
\newblock
  \bibinfo{howpublished}{\url{http://www.pentest-standard.org/index.php/PTES_Technical_Guidelines}}.
\newblock
\newblock
\shownote{Accessed: 2022-09-30}.


\bibitem[sql({[n.\,d.]})]%
        {sqlmap}
 \bibinfo{year}{[n.\,d.]}\natexlab{}.
\newblock \bibinfo{title}{sqlmap: automatic SQL injection and database takeover
  tool}.
\newblock \bibinfo{howpublished}{\url{https://sqlmap.org/}}.
\newblock
\newblock
\shownote{Accessed: 2022-09-30}.


\bibitem[pri({[n.\,d.]})]%
        {printernightmare}
 \bibinfo{year}{[n.\,d.]}\natexlab{}.
\newblock \bibinfo{title}{Windows Print Spooler Remote Code Execution
  Vulnerability (CVE-2021-34527)}.
\newblock
  \bibinfo{howpublished}{\url{https://msrc.microsoft.com/update-guide/vulnerability/CVE-2021-34527}}.
\newblock
\newblock
\shownote{Accessed: 2022-09-30}.


\bibitem[zdi({[n.\,d.]})]%
        {zdi}
 \bibinfo{year}{[n.\,d.]}\natexlab{}.
\newblock \bibinfo{title}{Zero Day Initiative}.
\newblock \bibinfo{howpublished}{\url{https://www.zerodayinitiative.com/blog}}.
\newblock
\newblock
\shownote{Accessed: 2022-09-30}.


\bibitem[nis(7 06)]%
        {nisg}
 \bibinfo{year}{2016-07-06}\natexlab{}.
\newblock \showarticletitle{DIRECTIVE (EU) 2016/1148 OF THE EUROPEAN PARLIAMENT
  AND OF THE COUNCIL of 6 July 2016 concerning measures for a high common level
  of security of network and information systems across the Union}.
\newblock
  \bibinfo{howpublished}{\url{https://eur-lex.europa.eu/legal-content/EN/TXT/PDF/?uri=CELEX:32016L1148}}.
\newblock \bibinfo{journal}{\emph{Official Journal of the European Union}}
  \bibinfo{volume}{L 194} (\bibinfo{year}{2016-07-06}), \bibinfo{pages}{1--30}.
\newblock


\bibitem[Aniche et~al\mbox{.}(2018)]%
        {aniche2018modern}
\bibfield{author}{\bibinfo{person}{Maur{\'\i}cio Aniche},
  \bibinfo{person}{Christoph Treude}, \bibinfo{person}{Igor Steinmacher},
  \bibinfo{person}{Igor Wiese}, \bibinfo{person}{Gustavo Pinto},
  \bibinfo{person}{Margaret-Anne Storey}, {and}
  \bibinfo{person}{Marco~Aur{\'e}lio Gerosa}.} \bibinfo{year}{2018}\natexlab{}.
\newblock \showarticletitle{How modern news aggregators help development
  communities shape and share knowledge}. In
  \bibinfo{booktitle}{\emph{Proceedings of the 40th International conference on
  software engineering}}. \bibinfo{pages}{499--510}.
\newblock


\bibitem[Bhuiyan et~al\mbox{.}(2020)]%
        {bhuiyan2020vulnerability}
\bibfield{author}{\bibinfo{person}{Farzana~Ahamed Bhuiyan},
  \bibinfo{person}{Akond Rahman}, {and} \bibinfo{person}{Patrick Morrison}.}
  \bibinfo{year}{2020}\natexlab{}.
\newblock \showarticletitle{Vulnerability discovery strategies used in software
  projects}. In \bibinfo{booktitle}{\emph{Proceedings of the 35th IEEE/ACM
  International Conference on Automated Software Engineering Workshops}}.
  \bibinfo{pages}{13--18}.
\newblock


\bibitem[Blankenship(1986)]%
        {thehackermanifesto}
\bibfield{author}{\bibinfo{person}{Loyd Blankenship}.}
  \bibinfo{year}{1986}\natexlab{}.
\newblock \showarticletitle{The Conscience of a Hacker}.
\newblock \bibinfo{journal}{\emph{Phrack}}  \bibinfo{volume}{7}
  (\bibinfo{date}{Jan.} \bibinfo{year}{1986}).
\newblock
\urldef\tempurl%
\url{http://www.phrack.org/archives/issues/7/3.txt}
\showURL{%
\tempurl}


\bibitem[Boyanov(2018)]%
        {boyanov2018educational}
\bibfield{author}{\bibinfo{person}{Petar Boyanov}.}
  \bibinfo{year}{2018}\natexlab{}.
\newblock \showarticletitle{Educational exploiting the information resources
  and invading the security mechanisms of the operating system Windows 7 with
  the exploit Eternalblue and Backdoor Doublepulsar}.
\newblock \bibinfo{journal}{\emph{Association Scientific and Applied Research}}
   \bibinfo{volume}{14} (\bibinfo{year}{2018}), \bibinfo{pages}{34}.
\newblock


\bibitem[Braun and Clarke(2019)]%
        {braun2019reflecting}
\bibfield{author}{\bibinfo{person}{Virginia Braun} {and}
  \bibinfo{person}{Victoria Clarke}.} \bibinfo{year}{2019}\natexlab{}.
\newblock \showarticletitle{Reflecting on reflexive thematic analysis}.
\newblock \bibinfo{journal}{\emph{Qualitative research in sport, exercise and
  health}} \bibinfo{volume}{11}, \bibinfo{number}{4} (\bibinfo{year}{2019}),
  \bibinfo{pages}{589--597}.
\newblock


\bibitem[Bukac et~al\mbox{.}(2014)]%
        {bukac2014red}
\bibfield{author}{\bibinfo{person}{Vit Bukac}, \bibinfo{person}{Vaclav Lorenc},
  {and} \bibinfo{person}{Vashek Maty{\'a}{\v{s}}}.}
  \bibinfo{year}{2014}\natexlab{}.
\newblock \showarticletitle{Red queen’s race: APT win-win game}. In
  \bibinfo{booktitle}{\emph{Cambridge International Workshop on Security
  Protocols}}. Springer, \bibinfo{pages}{55--61}.
\newblock


\bibitem[Ceccato et~al\mbox{.}(2019)]%
        {ceccato2019understanding}
\bibfield{author}{\bibinfo{person}{Mariano Ceccato}, \bibinfo{person}{Paolo
  Tonella}, \bibinfo{person}{Cataldo Basile}, \bibinfo{person}{Paolo Falcarin},
  \bibinfo{person}{Marco Torchiano}, \bibinfo{person}{Bart Coppens}, {and}
  \bibinfo{person}{Bjorn De~Sutter}.} \bibinfo{year}{2019}\natexlab{}.
\newblock \showarticletitle{Understanding the behaviour of hackers while
  performing attack tasks in a professional setting and in a public challenge}.
\newblock \bibinfo{journal}{\emph{Empirical Software Engineering}}
  \bibinfo{volume}{24} (\bibinfo{year}{2019}), \bibinfo{pages}{240--286}.
\newblock


\bibitem[Durumeric et~al\mbox{.}(2014)]%
        {durumeric2014matter}
\bibfield{author}{\bibinfo{person}{Zakir Durumeric}, \bibinfo{person}{Frank
  Li}, \bibinfo{person}{James Kasten}, \bibinfo{person}{Johanna Amann},
  \bibinfo{person}{Jethro Beekman}, \bibinfo{person}{Mathias Payer},
  \bibinfo{person}{Nicolas Weaver}, \bibinfo{person}{David Adrian},
  \bibinfo{person}{Vern Paxson}, \bibinfo{person}{Michael Bailey},
  {et~al\mbox{.}}} \bibinfo{year}{2014}\natexlab{}.
\newblock \showarticletitle{The matter of heartbleed}. In
  \bibinfo{booktitle}{\emph{Proceedings of the 2014 conference on internet
  measurement conference}}. \bibinfo{pages}{475--488}.
\newblock


\bibitem[Ferretti et~al\mbox{.}(2016)]%
        {ferretti2016automatic}
\bibfield{author}{\bibinfo{person}{Stefano Ferretti}, \bibinfo{person}{Silvia
  Mirri}, \bibinfo{person}{Catia Prandi}, {and} \bibinfo{person}{Paola
  Salomoni}.} \bibinfo{year}{2016}\natexlab{}.
\newblock \showarticletitle{Automatic web content personalization through
  reinforcement learning}.
\newblock \bibinfo{journal}{\emph{Journal of Systems and Software}}
  \bibinfo{volume}{121} (\bibinfo{year}{2016}), \bibinfo{pages}{157--169}.
\newblock


\bibitem[Francis et~al\mbox{.}(2010)]%
        {francis2010adequate}
\bibfield{author}{\bibinfo{person}{Jill~J Francis}, \bibinfo{person}{Marie
  Johnston}, \bibinfo{person}{Clare Robertson}, \bibinfo{person}{Liz
  Glidewell}, \bibinfo{person}{Vikki Entwistle}, \bibinfo{person}{Martin~P
  Eccles}, {and} \bibinfo{person}{Jeremy~M Grimshaw}.}
  \bibinfo{year}{2010}\natexlab{}.
\newblock \showarticletitle{What is an adequate sample size? Operationalising
  data saturation for theory-based interview studies}.
\newblock \bibinfo{journal}{\emph{Psychology and health}} \bibinfo{volume}{25},
  \bibinfo{number}{10} (\bibinfo{year}{2010}), \bibinfo{pages}{1229--1245}.
\newblock


\bibitem[Gascon et~al\mbox{.}(2015)]%
        {gascon2015pulsar}
\bibfield{author}{\bibinfo{person}{Hugo Gascon}, \bibinfo{person}{Christian
  Wressnegger}, \bibinfo{person}{Fabian Yamaguchi}, \bibinfo{person}{Daniel
  Arp}, {and} \bibinfo{person}{Konrad Rieck}.} \bibinfo{year}{2015}\natexlab{}.
\newblock \showarticletitle{Pulsar: Stateful black-box fuzzing of proprietary
  network protocols}. In \bibinfo{booktitle}{\emph{Security and Privacy in
  Communication Networks: 11th EAI International Conference, SecureComm 2015,
  Dallas, TX, USA, October 26-29, 2015, Proceedings 11}}. Springer,
  \bibinfo{pages}{330--347}.
\newblock


\bibitem[Guba et~al\mbox{.}(1994)]%
        {guba1994competing}
\bibfield{author}{\bibinfo{person}{Egon~G Guba}, \bibinfo{person}{Yvonna~S
  Lincoln}, {et~al\mbox{.}}} \bibinfo{year}{1994}\natexlab{}.
\newblock \showarticletitle{Competing paradigms in qualitative research}.
\newblock \bibinfo{journal}{\emph{Handbook of qualitative research}}
  \bibinfo{volume}{2}, \bibinfo{number}{163-194} (\bibinfo{year}{1994}),
  \bibinfo{pages}{105}.
\newblock


\bibitem[Guest et~al\mbox{.}(2006)]%
        {guest2006many}
\bibfield{author}{\bibinfo{person}{Greg Guest}, \bibinfo{person}{Arwen Bunce},
  {and} \bibinfo{person}{Laura Johnson}.} \bibinfo{year}{2006}\natexlab{}.
\newblock \showarticletitle{How many interviews are enough? An experiment with
  data saturation and variability}.
\newblock \bibinfo{journal}{\emph{Field methods}} \bibinfo{volume}{18},
  \bibinfo{number}{1} (\bibinfo{year}{2006}), \bibinfo{pages}{59--82}.
\newblock


\bibitem[Guzman({[n.\,d.]})]%
        {owasp_firmware}
\bibfield{author}{\bibinfo{person}{Aaron Guzman}.}
  \bibinfo{year}{[n.\,d.]}\natexlab{}.
\newblock \bibinfo{title}{OWASP Firmware Security Testing Methodology}.
\newblock
  \bibinfo{howpublished}{\url{https://scriptingxss.gitbook.io/firmware-security-testing-methodology/}}.
\newblock
\newblock
\shownote{Accessed: 2022-09-30}.


\bibitem[Guzman and Bassem(2020)]%
        {owasp_isvs}
\bibfield{author}{\bibinfo{person}{Aaron Guzman} {and} \bibinfo{person}{Cedric
  Bassem}.} \bibinfo{year}{2020}\natexlab{}.
\newblock \bibinfo{title}{OWASP IoT Security Verification Standard}.
\newblock
  \bibinfo{howpublished}{\url{https://github.com/OWASP/IoT-Security-Verification-Standard-ISVS/releases/download/1.0RC/OWASP_ISVS-1.0RC-en_WIP_.pdf}}.
\newblock


\bibitem[Happe and Jürgen(2023)]%
        {getting_pwned}
\bibfield{author}{\bibinfo{person}{Andreas Happe} {and} \bibinfo{person}{Cito
  Jürgen}.} \bibinfo{year}{2023}\natexlab{}.
\newblock \showarticletitle{Getting pwn’d by AI: Penetration Testing with
  Large Language Models}. In \bibinfo{booktitle}{\emph{Proceedings of the 31st
  ACM Joint European Software Engineering Conference and Symposium on the
  Foundations of Software Engineering}} (San Francisco, USA)
  \emph{(\bibinfo{series}{ESEC/FSE 2023})}. \bibinfo{publisher}{Association for
  Computing Machinery}, \bibinfo{address}{New York, NY, USA},
  \bibinfo{numpages}{5}~pages.
\newblock
\urldef\tempurl%
\url{https://doi.org/10.1145/3611643.3613083}
\showDOI{\tempurl}


\bibitem[Harang and Ducau(2018)]%
        {harang2018measuring}
\bibfield{author}{\bibinfo{person}{Richard Harang} {and}
  \bibinfo{person}{Felipe~N Ducau}.} \bibinfo{year}{2018}\natexlab{}.
\newblock \showarticletitle{Measuring the speed of the Red Queen’s Race}.
\newblock \bibinfo{journal}{\emph{BlackHat: Las Vegas, NV, USA}}
  (\bibinfo{year}{2018}).
\newblock


\bibitem[Holguera et~al\mbox{.}(2022)]%
        {owasp_masvs}
\bibfield{author}{\bibinfo{person}{Carlos Holguera}, \bibinfo{person}{Bernhard
  Müller}, \bibinfo{person}{Sven Schleier}, {and} \bibinfo{person}{Jeroen
  Willemsen}.} \bibinfo{year}{2022}\natexlab{}.
\newblock \bibinfo{title}{OWASP Mobile Application Security Verification
  Standard}.
\newblock
  \bibinfo{howpublished}{\url{https://github.com/OWASP/owasp-masvs/releases/latest/download/OWASP_MASVS-v1.4.2-en.pdf}}.
\newblock


\bibitem[Huaman et~al\mbox{.}(2021)]%
        {huaman2021large}
\bibfield{author}{\bibinfo{person}{Nicolas Huaman}, \bibinfo{person}{Bennet von
  Skarczinski}, \bibinfo{person}{Dominik Wermke}, \bibinfo{person}{Christian
  Stransky}, \bibinfo{person}{Yasemin Acar}, \bibinfo{person}{Arne
  Drei{\ss}igacker}, {and} \bibinfo{person}{Sascha Fahl}.}
  \bibinfo{year}{2021}\natexlab{}.
\newblock \showarticletitle{A large-scale interview study on information
  security in and attacks against small and medium-sized enterprises}. In
  \bibinfo{booktitle}{\emph{In 30th USENIX Security Symposium}}.
\newblock


\bibitem[(ISC)2(2022)]%
        {isc2}
\bibfield{author}{\bibinfo{person}{(ISC)2}.} \bibinfo{year}{2022}\natexlab{}.
\newblock \bibinfo{title}{(ISC)2 CYBERSECURITY WORKFORCE STUDY 2022}.
\newblock
  \bibinfo{howpublished}{\url{https://www.isc2.org//-/media/ISC2/Research/2022-WorkForce-Study/ISC2-Cybersecurity-Workforce-Study.ashx}}.
\newblock
\newblock
\shownote{Accessed: 2023-04-28}.


\bibitem[Katakis et~al\mbox{.}(2009)]%
        {katakis2009adaptive}
\bibfield{author}{\bibinfo{person}{Ioannis Katakis}, \bibinfo{person}{Grigorios
  Tsoumakas}, \bibinfo{person}{Evangelos Banos}, \bibinfo{person}{Nick
  Bassiliades}, {and} \bibinfo{person}{Ioannis Vlahavas}.}
  \bibinfo{year}{2009}\natexlab{}.
\newblock \showarticletitle{An adaptive personalized news dissemination
  system}.
\newblock \bibinfo{journal}{\emph{Journal of intelligent information systems}}
  \bibinfo{volume}{32} (\bibinfo{year}{2009}), \bibinfo{pages}{191--212}.
\newblock


\bibitem[Kettle(2019)]%
        {http_smuggling}
\bibfield{author}{\bibinfo{person}{James Kettle}.}
  \bibinfo{year}{2019}\natexlab{}.
\newblock \bibinfo{title}{HTTP Desync Attacks: Request Smuggling Reborn}.
\newblock
  \bibinfo{howpublished}{\url{https://portswigger.net/research/http-desync-attacks-request-smuggling-reborn}}.
\newblock
\newblock
\shownote{Accessed: 2023-08-18}.


\bibitem[Kettle(2022)]%
        {http_smuggling_2}
\bibfield{author}{\bibinfo{person}{James Kettle}.}
  \bibinfo{year}{2022}\natexlab{}.
\newblock \bibinfo{title}{Browser-Powered Desync Attacks: A New Frontier in
  HTTP Request Smuggling}.
\newblock
  \bibinfo{howpublished}{\url{https://portswigger.net/research/browser-powered-desync-attacks}}.
\newblock
\newblock
\shownote{Accessed: 2023-08-18}.


\bibitem[Kotulic and Clark(2004)]%
        {kotulic2004there}
\bibfield{author}{\bibinfo{person}{Andrew~G Kotulic} {and}
  \bibinfo{person}{Jan~Guynes Clark}.} \bibinfo{year}{2004}\natexlab{}.
\newblock \showarticletitle{Why there aren’t more information security
  research studies}.
\newblock \bibinfo{journal}{\emph{Information \& Management}}
  \bibinfo{volume}{41}, \bibinfo{number}{5} (\bibinfo{year}{2004}),
  \bibinfo{pages}{597--607}.
\newblock


\bibitem[Lake(2022)]%
        {lackofsecuritypersonell}
\bibfield{author}{\bibinfo{person}{Sydney Lake}.}
  \bibinfo{year}{2022}\natexlab{}.
\newblock \bibinfo{title}{The cybersecurity industry is short 3.4 million
  workers---that’s good news for cyber wages}.
\newblock
  \bibinfo{howpublished}{\url{https://fortune.com/education/articles/the-cybersecurity-industry-is-short-3-4-million-workers-thats-good-news-for-cyber-wages/}}.
\newblock
\newblock
\shownote{Accessed: 2023-04-28}.


\bibitem[Mackenzie and Knipe(2006)]%
        {mackenzie2006research}
\bibfield{author}{\bibinfo{person}{Noella Mackenzie} {and}
  \bibinfo{person}{Sally Knipe}.} \bibinfo{year}{2006}\natexlab{}.
\newblock \showarticletitle{Research dilemmas: Paradigms, methods and
  methodology.}
\newblock \bibinfo{journal}{\emph{Issues in educational research}}
  \bibinfo{volume}{16}, \bibinfo{number}{2} (\bibinfo{year}{2006}),
  \bibinfo{pages}{193--205}.
\newblock


\bibitem[Munaiah et~al\mbox{.}(2019)]%
        {munaiah2019characterizing}
\bibfield{author}{\bibinfo{person}{Nuthan Munaiah}, \bibinfo{person}{Akond
  Rahman}, \bibinfo{person}{Justin Pelletier}, \bibinfo{person}{Laurie
  Williams}, {and} \bibinfo{person}{Andrew Meneely}.}
  \bibinfo{year}{2019}\natexlab{}.
\newblock \showarticletitle{Characterizing attacker behavior in a cybersecurity
  penetration testing competition}. In \bibinfo{booktitle}{\emph{2019 ACM/IEEE
  International Symposium on Empirical Software Engineering and Measurement
  (ESEM)}}. IEEE, \bibinfo{pages}{1--6}.
\newblock


\bibitem[Potter and McGraw(2004)]%
        {defensive3}
\bibfield{author}{\bibinfo{person}{Bruce Potter} {and} \bibinfo{person}{Gary
  McGraw}.} \bibinfo{year}{2004}\natexlab{}.
\newblock \showarticletitle{Software security testing}.
\newblock \bibinfo{journal}{\emph{IEEE Security \& Privacy}}
  \bibinfo{volume}{2}, \bibinfo{number}{5} (\bibinfo{year}{2004}),
  \bibinfo{pages}{81--85}.
\newblock


\bibitem[Saad and Mitchell(2020)]%
        {owasp_otg}
\bibfield{author}{\bibinfo{person}{Elie Saad} {and} \bibinfo{person}{Rick
  Mitchell}.} \bibinfo{year}{2020}\natexlab{}.
\newblock \bibinfo{title}{OWASP Web Security Testing Guide}.
\newblock
  \bibinfo{howpublished}{\url{https://github.com/OWASP/wstg/releases/download/v4.2/wstg-v4.2.pdf}}.
\newblock


\bibitem[Saunders and Tosey(2013)]%
        {saunders2013layers}
\bibfield{author}{\bibinfo{person}{MNK Saunders} {and} \bibinfo{person}{PC
  Tosey}.} \bibinfo{year}{2013}\natexlab{}.
\newblock \bibinfo{booktitle}{\emph{The layers of research design}}.
\newblock \bibinfo{type}{{T}echnical {R}eport}.
  \bibinfo{institution}{University of Surrey}.
\newblock


\bibitem[Schleier et~al\mbox{.}(2022)]%
        {owasp_mtg}
\bibfield{author}{\bibinfo{person}{Sven Schleier}, \bibinfo{person}{Bernhard
  Mueller}, \bibinfo{person}{Carlos Holguera}, {and} \bibinfo{person}{Jeroen
  Willemsen}.} \bibinfo{year}{2022}\natexlab{}.
\newblock \bibinfo{title}{OWASP Mobile Application Security Testing Guide}.
\newblock
  \bibinfo{howpublished}{\url{https://github.com/OWASP/owasp-mastg/releases/latest/download/OWASP_MASTG-v1.5.0.pdf}}.
\newblock


\bibitem[Singer et~al\mbox{.}(2014)]%
        {singer2014software}
\bibfield{author}{\bibinfo{person}{Leif Singer}, \bibinfo{person}{Fernando
  Figueira~Filho}, {and} \bibinfo{person}{Margaret-Anne Storey}.}
  \bibinfo{year}{2014}\natexlab{}.
\newblock \showarticletitle{Software engineering at the speed of light: how
  developers stay current using twitter}. In
  \bibinfo{booktitle}{\emph{Proceedings of the 36th International Conference on
  Software Engineering}}. \bibinfo{pages}{211--221}.
\newblock


\bibitem[Smith et~al\mbox{.}(2020)]%
        {smith2020case}
\bibfield{author}{\bibinfo{person}{Justin Smith}, \bibinfo{person}{Christopher
  Theisen}, {and} \bibinfo{person}{Titus Barik}.}
  \bibinfo{year}{2020}\natexlab{}.
\newblock \showarticletitle{A Case Study of Software Security Red Teams at
  Microsoft}. In \bibinfo{booktitle}{\emph{2020 IEEE Symposium on Visual
  Languages and Human-Centric Computing (VL/HCC)}}. IEEE,
  \bibinfo{pages}{1--10}.
\newblock


\bibitem[Strom et~al\mbox{.}(2018)]%
        {strom2018mitre}
\bibfield{author}{\bibinfo{person}{Blake~E Strom}, \bibinfo{person}{Andy
  Applebaum}, \bibinfo{person}{Doug~P Miller}, \bibinfo{person}{Kathryn~C
  Nickels}, \bibinfo{person}{Adam~G Pennington}, {and} \bibinfo{person}{Cody~B
  Thomas}.} \bibinfo{year}{2018}\natexlab{}.
\newblock \showarticletitle{Mitre att\&ck: Design and philosophy}.
\newblock In \bibinfo{booktitle}{\emph{Technical report}}.
  \bibinfo{publisher}{The MITRE Corporation}.
\newblock


\bibitem[Summers(2015)]%
        {summers2015hackers}
\bibfield{author}{\bibinfo{person}{Timothy~C Summers}.}
  \bibinfo{year}{2015}\natexlab{}.
\newblock \bibinfo{booktitle}{\emph{How hackers think: A mixed method study of
  mental models and cognitive patterns of high-tech wizards}}.
\newblock \bibinfo{publisher}{Case Western Reserve University}.
\newblock


\bibitem[Takanen et~al\mbox{.}(2018)]%
        {defensive4}
\bibfield{author}{\bibinfo{person}{Ari Takanen}, \bibinfo{person}{Jared~D
  Demott}, \bibinfo{person}{Charles Miller}, {and} \bibinfo{person}{Atte
  Kettunen}.} \bibinfo{year}{2018}\natexlab{}.
\newblock \bibinfo{booktitle}{\emph{Fuzzing for software security testing and
  quality assurance}}.
\newblock \bibinfo{publisher}{Artech House}.
\newblock


\bibitem[Torres et~al\mbox{.}(2011)]%
        {torres2011improving}
\bibfield{author}{\bibinfo{person}{Romina Torres}, \bibinfo{person}{Boris
  Tapia}, {et~al\mbox{.}}} \bibinfo{year}{2011}\natexlab{}.
\newblock \showarticletitle{Improving web api discovery by leveraging social
  information}. In \bibinfo{booktitle}{\emph{2011 IEEE International Conference
  on Web Services}}. IEEE, \bibinfo{pages}{744--745}.
\newblock


\bibitem[Trubiani et~al\mbox{.}(2019)]%
        {trubiani2019performance}
\bibfield{author}{\bibinfo{person}{Catia Trubiani}, \bibinfo{person}{Pooyan
  Jamshidi}, \bibinfo{person}{Jurgen Cito}, \bibinfo{person}{Weiyi Shang},
  \bibinfo{person}{Zhen~Ming Jiang}, {and} \bibinfo{person}{Markus Borg}.}
  \bibinfo{year}{2019}\natexlab{}.
\newblock \showarticletitle{Performance Issues? Hey DevOps, Mind the
  Uncertainty}.
\newblock \bibinfo{journal}{\emph{IEEE Software}} \bibinfo{volume}{36},
  \bibinfo{number}{02} (\bibinfo{year}{2019}), \bibinfo{pages}{110--117}.
\newblock


\bibitem[van~den Hout(2019)]%
        {phdthesis}
\bibfield{author}{\bibinfo{person}{Niek~Jan van~den Hout}.}
  \bibinfo{year}{2019}\natexlab{}.
\newblock \emph{\bibinfo{title}{Standardised Penetration Testing? Examining the
  Usefulness of Current Penetration Testing Methodologies}}.
\newblock \bibinfo{thesistype}{Ph.\,D. Dissertation}.
\newblock


\bibitem[van~der Stork et~al\mbox{.}(2021a)]%
        {owasp_top10}
\bibfield{author}{\bibinfo{person}{Andrew van~der Stork},
  \bibinfo{person}{Brian Glas}, \bibinfo{person}{Neil Smithline}, {and}
  \bibinfo{person}{Torsten Gigler}.} \bibinfo{year}{2021}\natexlab{a}.
\newblock \bibinfo{title}{OWASP Top 10:2021}.
\newblock \bibinfo{howpublished}{\url{https://owasp.org/Top10/0x00-notice/}}.
\newblock


\bibitem[van~der Stork et~al\mbox{.}(2021b)]%
        {owasp_asvs}
\bibfield{author}{\bibinfo{person}{Andrew van~der Stork}, \bibinfo{person}{Josh
  Grossman}, \bibinfo{person}{Daniel Cuthbert}, \bibinfo{person}{Elar Lang},
  {and} \bibinfo{person}{Jim Manico}.} \bibinfo{year}{2021}\natexlab{b}.
\newblock \bibinfo{title}{OWASP Application Security Verification Standard}.
\newblock
  \bibinfo{howpublished}{\url{https://github.com/OWASP/ASVS/raw/v4.0.3/4.0/OWASP+Application+Security+Verification+Standard+4.0.3-en.pdf}}.
\newblock


\bibitem[Wysopal et~al\mbox{.}(2006)]%
        {defensive1}
\bibfield{author}{\bibinfo{person}{Chris Wysopal}, \bibinfo{person}{Lucas
  Nelson}, \bibinfo{person}{Elfriede Dustin}, {and} \bibinfo{person}{Dino
  Dai~Zovi}.} \bibinfo{year}{2006}\natexlab{}.
\newblock \bibinfo{booktitle}{\emph{The art of software security testing:
  identifying software security flaws}}.
\newblock \bibinfo{publisher}{Pearson Education}.
\newblock


\bibitem[Xu et~al\mbox{.}(2018)]%
        {xu2018deeptype}
\bibfield{author}{\bibinfo{person}{Mengwei Xu}, \bibinfo{person}{Feng Qian},
  \bibinfo{person}{Qiaozhu Mei}, \bibinfo{person}{Kang Huang}, {and}
  \bibinfo{person}{Xuanzhe Liu}.} \bibinfo{year}{2018}\natexlab{}.
\newblock \showarticletitle{Deeptype: On-device deep learning for input
  personalization service with minimal privacy concern}.
\newblock \bibinfo{journal}{\emph{Proceedings of the ACM on Interactive,
  Mobile, Wearable and Ubiquitous Technologies}} \bibinfo{volume}{2},
  \bibinfo{number}{4} (\bibinfo{year}{2018}), \bibinfo{pages}{1--26}.
\newblock


\bibitem[Yessenov et~al\mbox{.}(2017)]%
        {yessenov2017demomatch}
\bibfield{author}{\bibinfo{person}{Kuat Yessenov}, \bibinfo{person}{Ivan
  Kuraj}, {and} \bibinfo{person}{Armando Solar-Lezama}.}
  \bibinfo{year}{2017}\natexlab{}.
\newblock \showarticletitle{DemoMatch: API discovery from demonstrations}.
\newblock \bibinfo{journal}{\emph{ACM SIGPLAN Notices}} \bibinfo{volume}{52},
  \bibinfo{number}{6} (\bibinfo{year}{2017}), \bibinfo{pages}{64--78}.
\newblock


\end{thebibliography}

\end{document}